\documentclass{iau} 
\usepackage{graphicx}

\title[Multiline observations of S255IR with ALMA] 
{Multiline observations of S255IR with ALMA}

\author[Igor Zinchenko, Sheng-Yuan Liu, et al.]   
{Igor Zinchenko$^1$,
 Sheng-Yuan Liu$^2$,
 Yu-Nung Su$^2$ \\ 
 \and 
 Petr Zemlyanukha$^1$}

\affiliation{$^1$Institute of Applied Physics of the Russian Academy of Sciences,\\
	   46 Uljanov~str., 603950 Nizhny Novgorod, Russia \\ email: {\tt zin@appl.sci-nnov.ru} \\[\affilskip]
$^2$Institute of Astronomy and Astrophysics, Academia Sinica,\\
P.O. Box 23-141, Taipei 10617, Taiwan, R.O.C \\ email: {\tt syliu@asiaa.sinica.edu.tw, ynsu@asiaa.sinica.edu.tw}}

\pubyear{2017}
\volume{332}  
\jname{Astrochemistry VII –-- Through the Cosmos from Galaxies to Planets}
\editors{Maria Cunningham, Tom Millar \& Yuri Aikawa, eds.}
\begin{document}

\maketitle

\begin{abstract}
We present preliminary results of the high  resolution $ (0.10^{\prime\prime} \times 0.15^{\prime\prime}) $ observations of the high mass star forming region S255IR with ALMA in several spectral windows from $ \sim 335 $~GHz to $ \sim 350 $~GHz. The main target lines were C$^{34}$S(7--6), CH$_3$CN($19_K-18_K$), CO(3--2) and SiO(8--7), however many other lines of various molecules have been detected, too. We present sample spectra and maps, discuss briefly the source structure and kinematics. A new, never predicted methanol maser line has been discovered.

\end{abstract}

\section{Introduction}
Disks around young high mass stars attract an enhanced attention nowadays. Their investigations are essential for understanding the process of high mass star formation, in particular the role of the disk accretion.

Here we present preliminary results of our recent observations of the S255IR-SMA1 core with ALMA. Our previous SMA observations \cite[(Zinchenko et al. 2012, 2015)]{Zin12,Zin15} as well as the data by \cite{Wang11} revealed a disk around the massive (20~M$_\odot$) YSO S255 NIRS3 associated with a spectacular outflow. A first disk-mediated accretion burst from a high mass YSO was detected recently here \cite[(Caratti O Garatti et al. 2017)]{Caratti16} following the methanol maser flare \cite[(Fujisawa et al. 2015)]{Fujisawa15}.

The goals of our ALMA observations included an investigation of the structure, kinematics and physical properties of this disk and associated outflow at a much higher angular resolution and sensitivity in comparison with the previous studies.

\section{Observations}
Observations of the S255IR region were carried out with ALMA in Band 7 at two epochs during Cycle 4. The first observing epoch on 2016 April 21, carried out with baselines ranging between 15 m and 612 m. The second epoch of observations was on 2016 September 9, with baselines ranging between 15 and 3143 m. Four spectral windows centered at around 335.4 GHz, 337.3 GHz, 349.0 GHz, and 346.6 GHz, with bandwidths of 1875.0 MHz, 234.4 MHz, 937.5 MHz, and 1875.0 MHz, respectively, were observed. 
Our primary target lines were the C$^{34}$S(7--6), CH$_3$CN($19_K-18_K$), CO(3--2) and SiO(8--7) transitions but many other lines are seen in the bands, too. The angular resolution of $ 0.10^{\prime\prime} \times 0.15^{\prime\prime} $ and sensitivity are much better than in our previous SMA data.

\section{Results}
In Fig.~\ref{fig:spectra} a part of the spectra is presented with possible line identifications (some of them need to be checked). The spectra are very rich, many complex molecules are observed, as expected for a hot core. A detailed analysis of the chemical composition will be given elsewhere. At the frequency of about 349.1~GHz a very bright apparently non-thermal maser line is detected.   Most probably this is the methanol $14_{1} - 14_{0}$ A$^{- +}$ transition. To the best of our knowledge this transition has never been observed or predicted to be masing. The maser emission is extended with a velocity gradient in same sense as the disk rotation. The peak of this maser emission is close to one of the clusters of the 6.7~GHz methanol masers observed by \cite{Moscadelli17}.

\begin{figure}
\centering
\includegraphics[width=\textwidth]{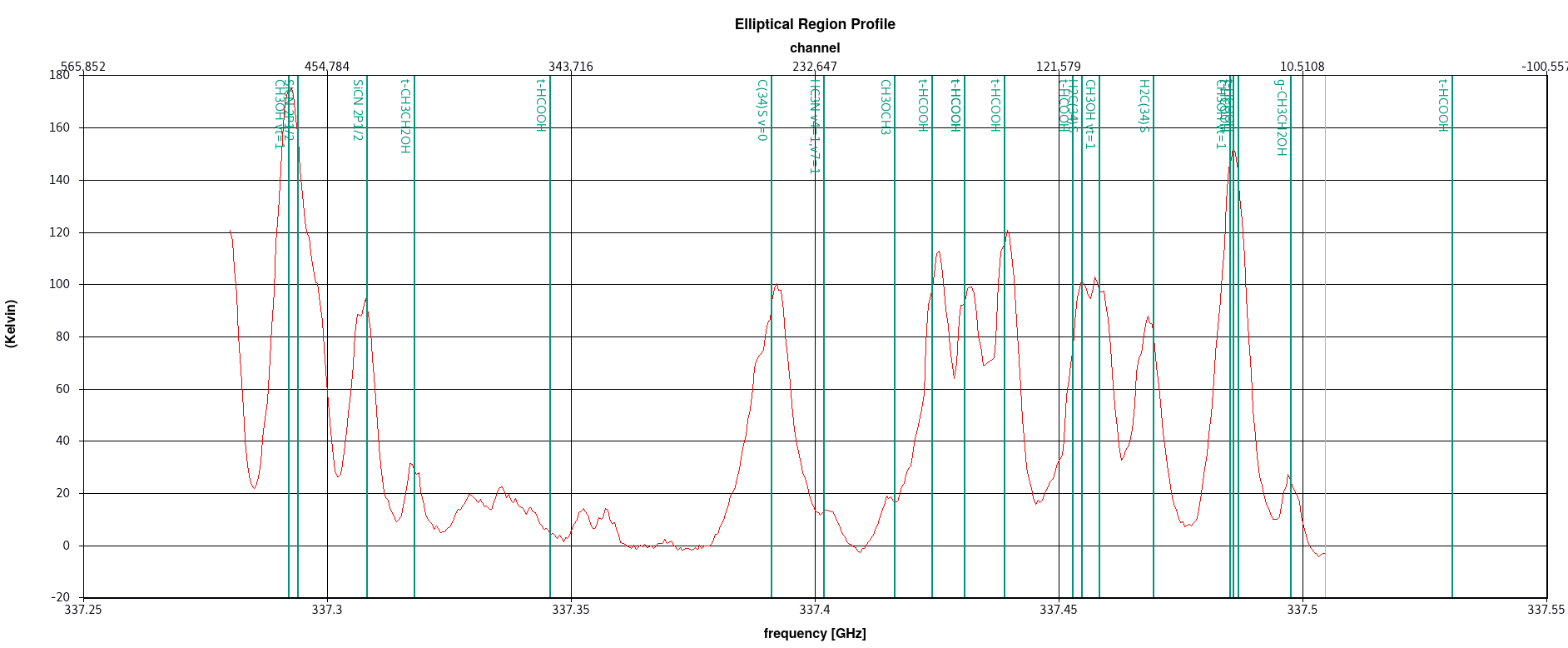}
\caption{Examples of the spectra measured in S255IR.}
\label{fig:spectra}
\end{figure}

\addtocounter{figure}{-1}
\begin{figure}
\centering
\includegraphics[width=\textwidth]{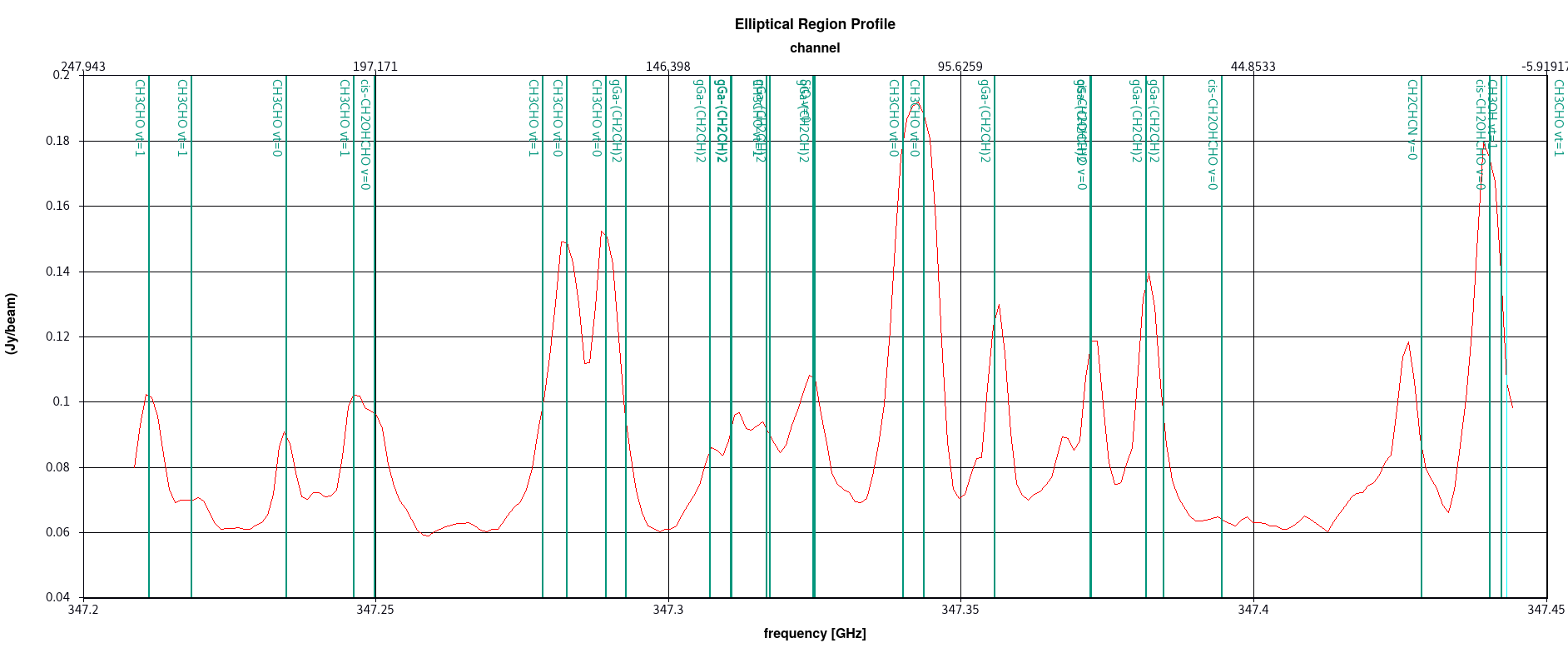}
\caption{continued}
\end{figure}

\addtocounter{figure}{-1}
\begin{figure}
\centering
\includegraphics[width=\textwidth]{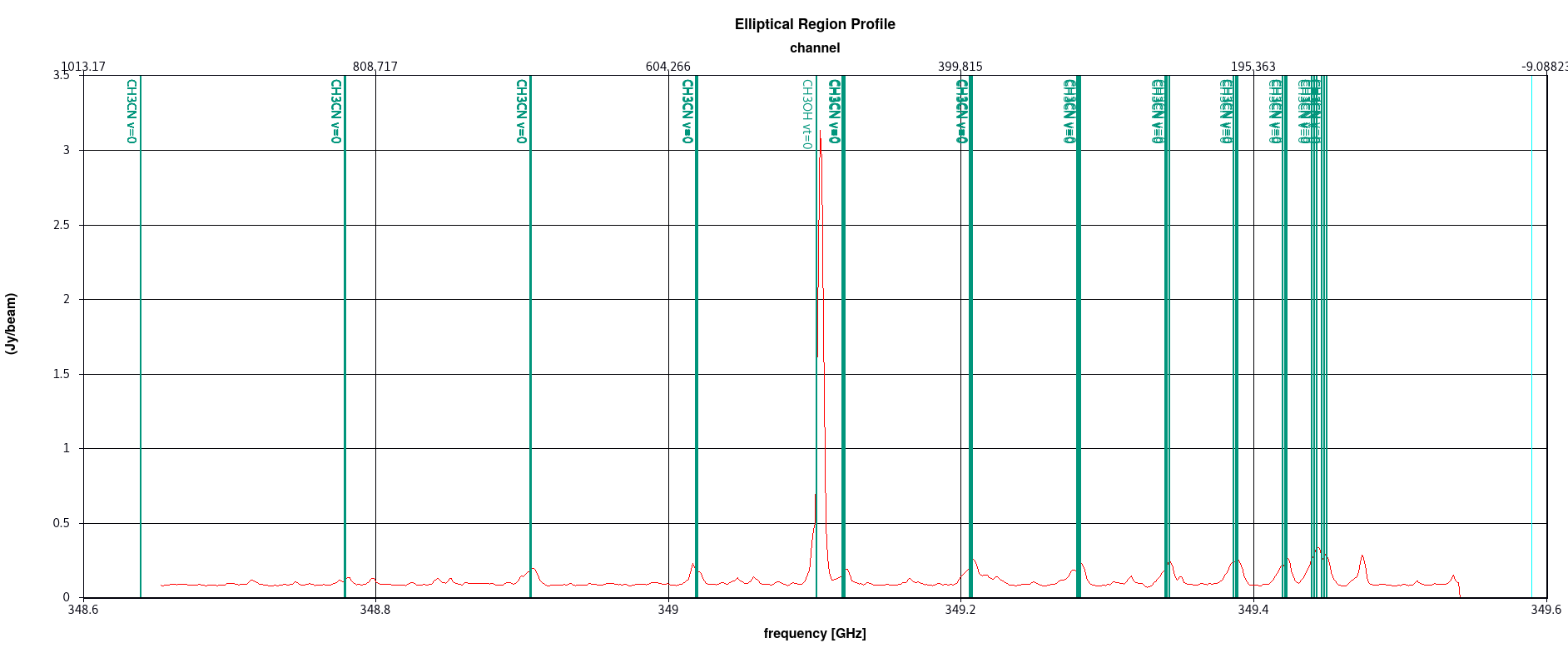}
\caption{continued}
\end{figure}

\begin{figure}
\begin{minipage}[b]{0.42\textwidth}
\includegraphics[width=\textwidth]{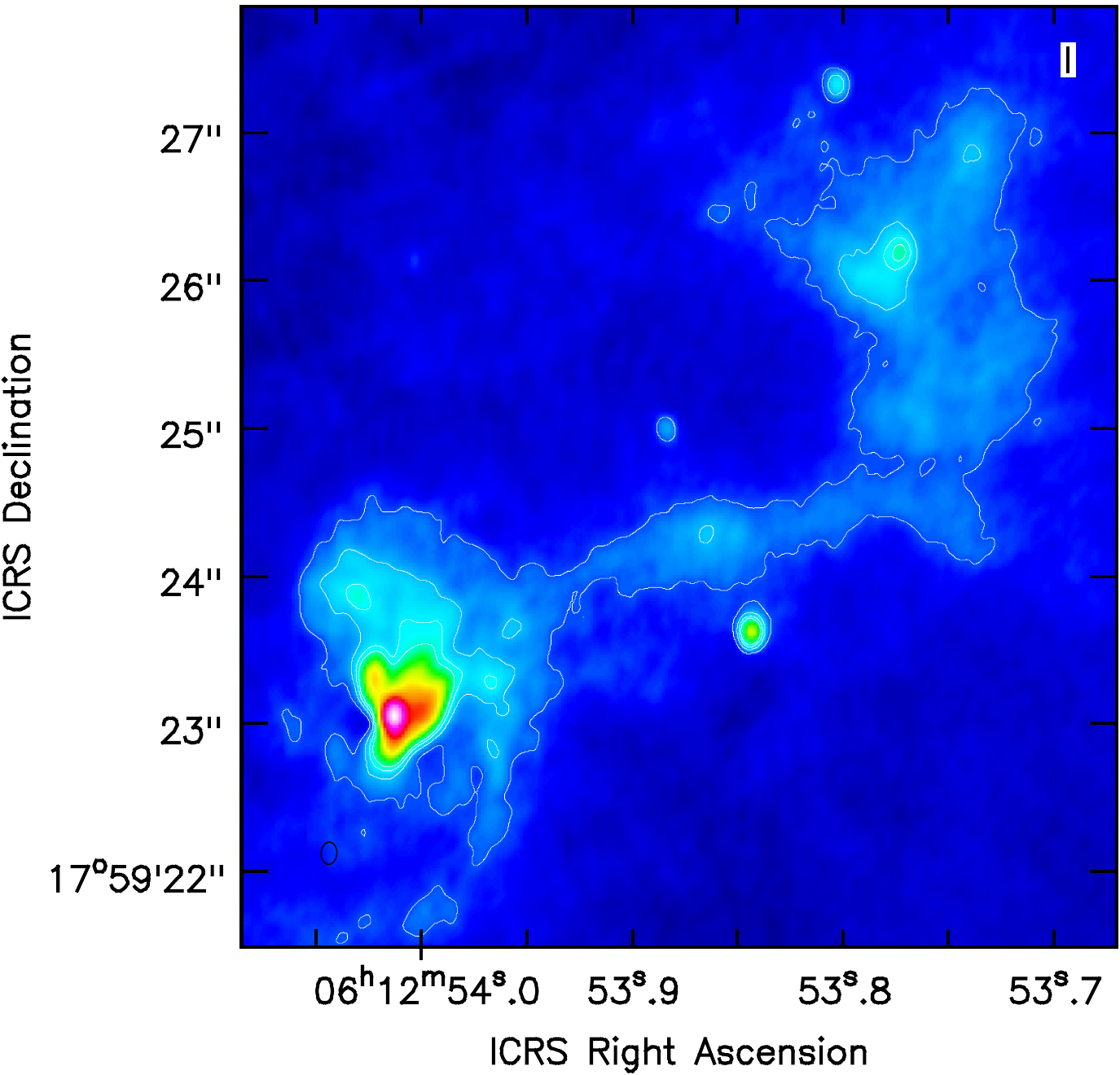}
\end{minipage}
\hfill
\begin{minipage}[b]{0.57\textwidth}
\includegraphics[width=\textwidth]{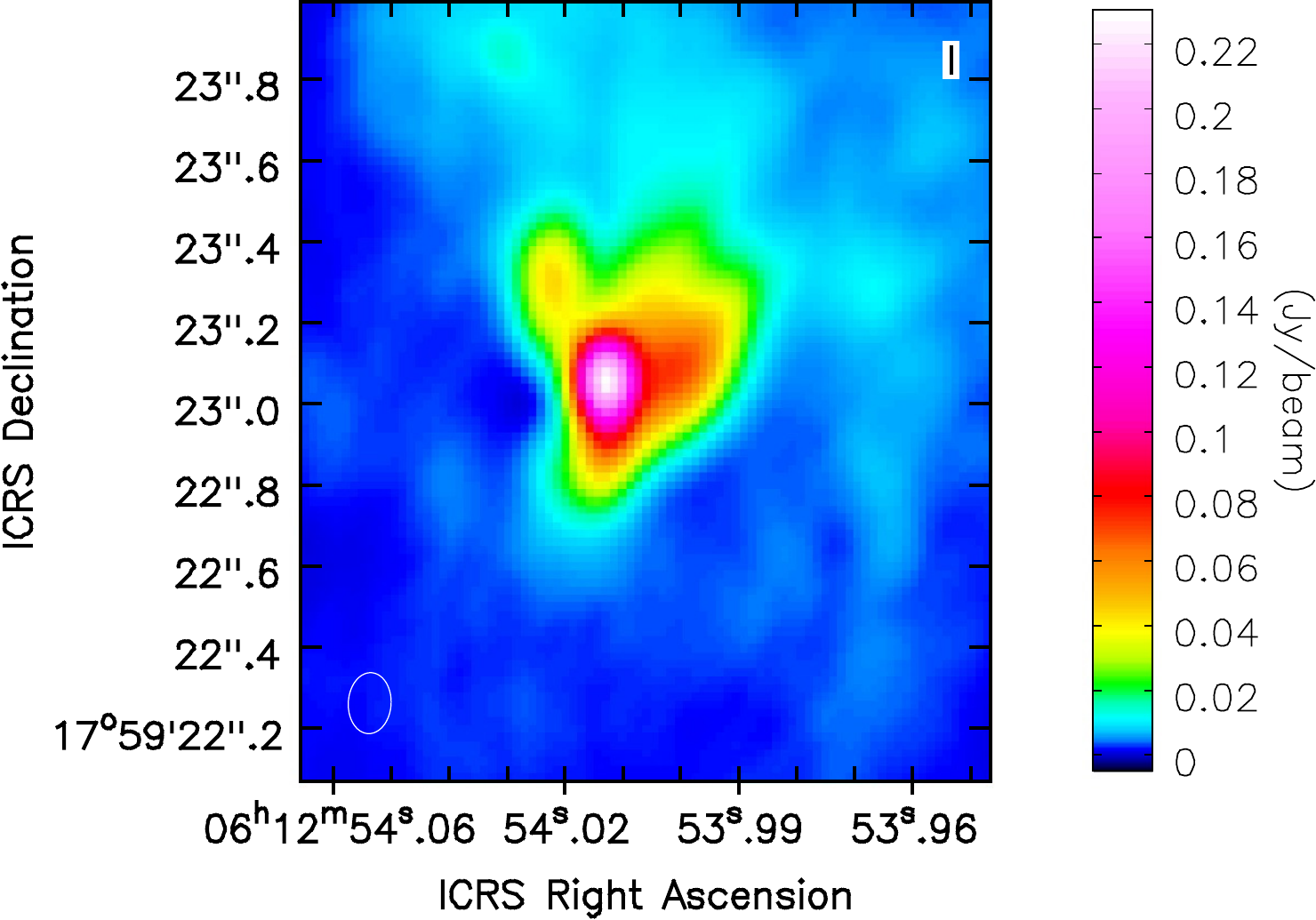}
\end{minipage}
\caption{The continuum maps at 0.85~mm of the whole S255IR area (left panel) and of the S255IR-SMA1 clump (right panel).}
\label{fig:cont}
\end{figure}

In Fig.~\ref{fig:cont} we present maps of the continuum emission of the S255IR area at 0.85~mm. They show in detail the main previously identified clumps: SMA1, SMA2 and SMA3 \cite[(Wang et al. 2011; Zinchenko et al. 2012, 2015)]{Wang11,Zin12,Zin15}. The SMA1 and SMA2 clumps appear to be physically related by a filamentary structure. The morphology of the SMA1 clump, which represents a disk around the massive YSO, looks rather complicated. 

\begin{figure}
\begin{minipage}[b]{0.54\textwidth}
\includegraphics[width=\textwidth]{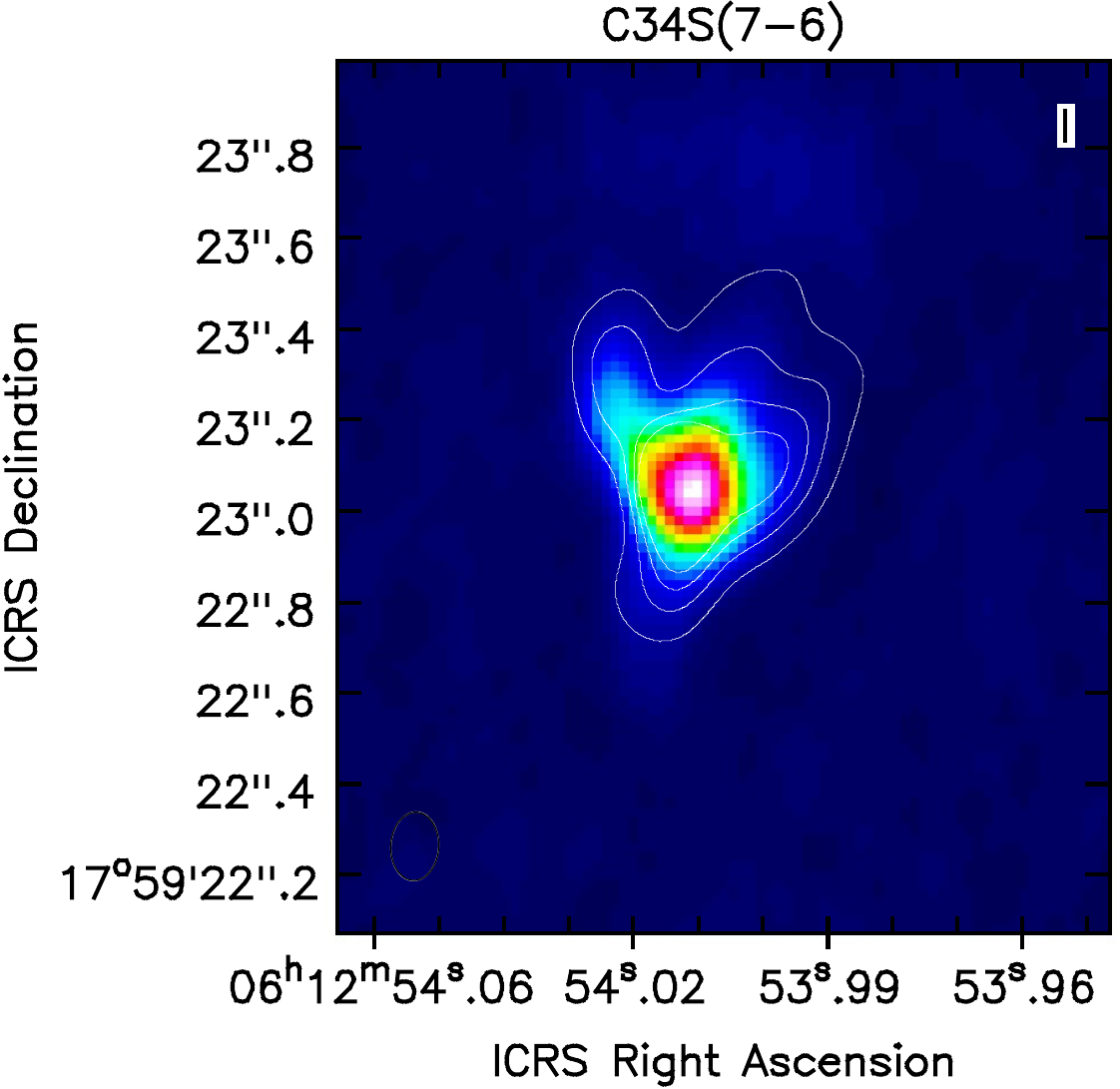}
\end{minipage}
\hfill
\begin{minipage}[b]{0.45\textwidth}
\includegraphics[width=\textwidth]{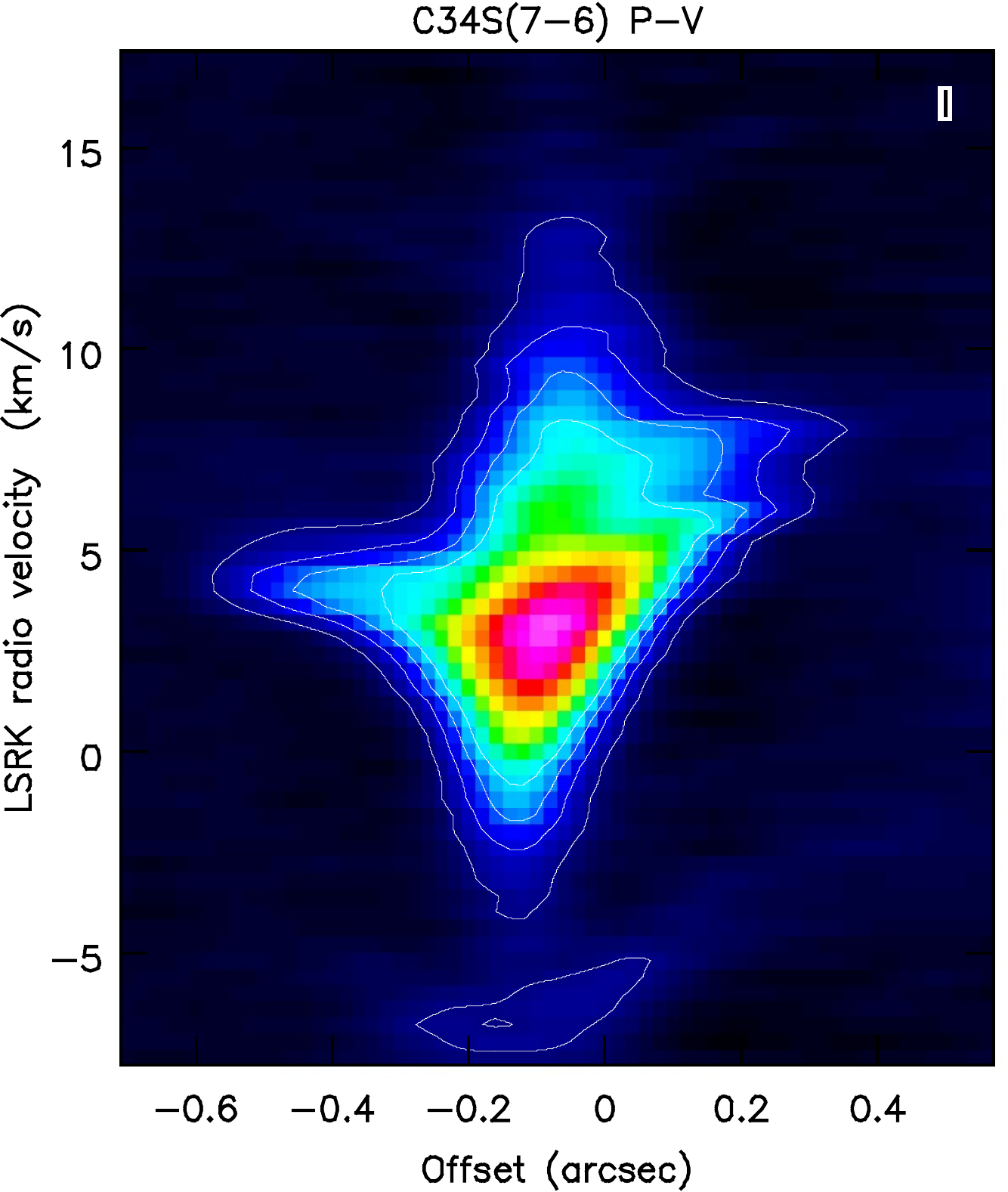}
\end{minipage}
\caption{Left panel: the map of the integrated C$^{34}$S(7--6) emission in the S255IR-SMA1 clump overlaid with the contours of the 0.85 mm continuum emission. Right panel: the position velocity diagram of the C$^{34}$S(7--6) emission along the major axis of the disk identified here earlier.}
\label{fig:c34s}
\end{figure}

In Fig.~\ref{fig:c34s} the map of the integrated intensity and the position-velocity diagram in the C$^{34}$S(7-6) line are presented. The latter  indicates a Keplerian-like rotation (seen in the other lines, too). 

\begin{figure}
\begin{minipage}[b]{0.48\textwidth}
\includegraphics[width=\textwidth]{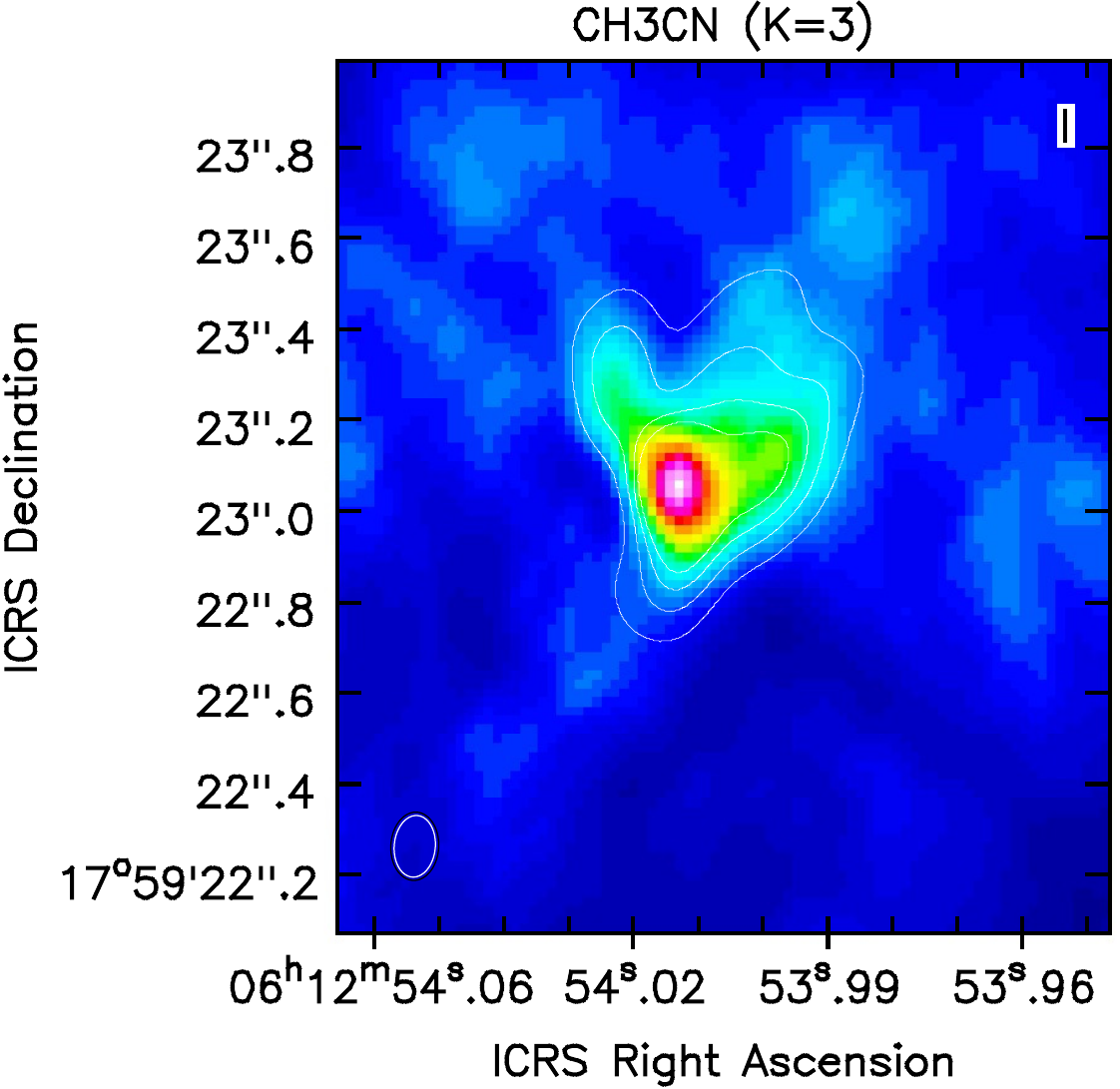}
\end{minipage}
\hfill
\begin{minipage}[b]{0.50\textwidth}
\includegraphics[width=\textwidth]{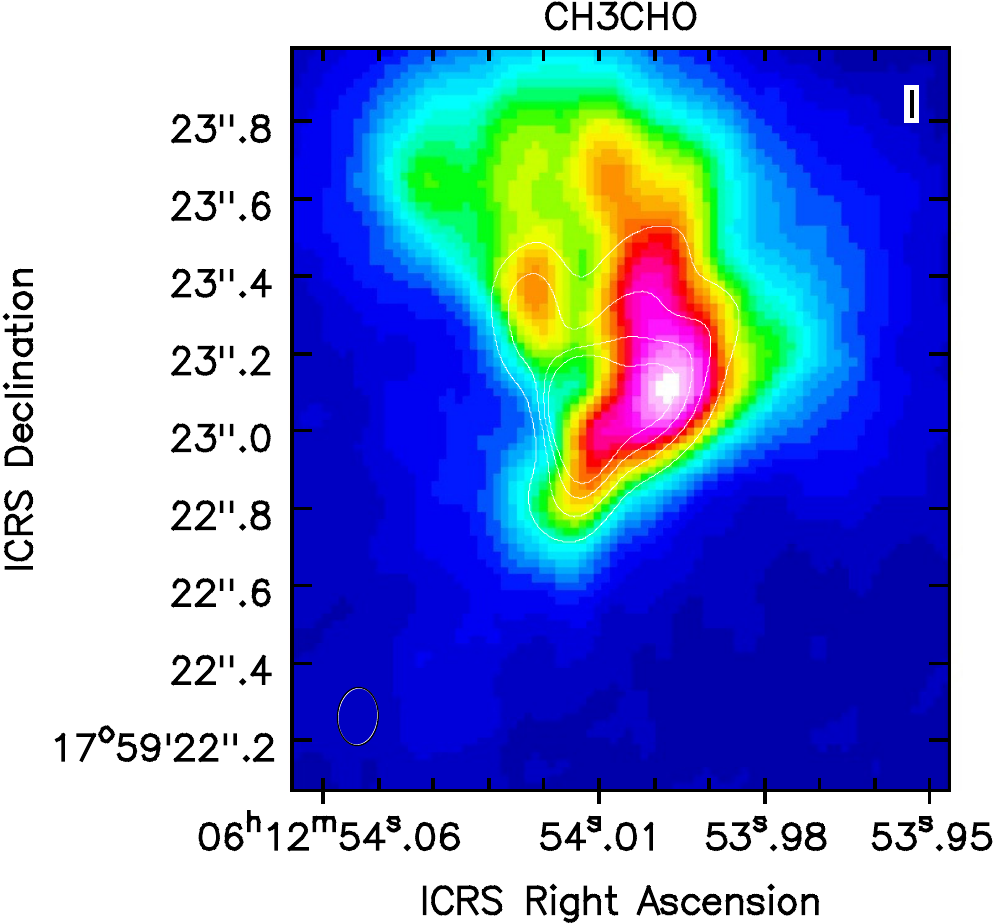}
\end{minipage}
\caption{Maps of the integrated intensity in the CH$_3$CN($19_3-18_3$) line (left panel) and in the CH$_3$CHO($18_5-17_5$) E line (right panel) overlaid with the continuum maps at 0.85~mm toward the S255IR-SMA1 clump.}
\label{fig:ch3cn}
\end{figure}

In Fig.~\ref{fig:ch3cn} we present sample maps for two other species: CH$_3$CN and CH$_3$CHO.

\section{Conclusions}
The preliminary results of our observations of the S255IR area with ALMA indicate a physical relation of the two major clumps. The disk around the massive YSO S255 NIRS3 has a complicated morphology and close to Keplerian rotation. The chemistry of this object is very rich. A new methanol maser line is detected.

This research was supported by the Russian Foundation for Basic Research (grant No. 15-02-06098) in the part of the preparation of the observations and preliminary data reduction, and by the Russian Science Foundation (grant No. 17-12-01256) in the part of the data analysis.

\end{document}